# Atomic optical amplifier with narrow bandwidth optical filtering

DUO PAN[1], TIANTIAN SHI[1], BIN LUO[2,*], JINGBIAO CHEN[1,**], HONG GUO[1]

[1]*State Key Laboratory of Advanced Optical Communicating, Systems and Networks, School of electronics Engineering and Computer Science, Peking University, Beijing 100871, China.*
[2]*State Key Laboratory of Information Photonics and Optical Communications , Beijing University of Posts and Telecommunications, Beijing 100876, China*
*Corresponding author: luobin@pku.edu.cn*
**Corresponding author: jbchen@pku.edu.cn*

**Taking advantages of ultra-narrow bandwidth and high noise rejection performance of the Faraday anomalous dispersion optical filter (FADOF), simultaneously with the coherent amplification of atomic stimulated emission, a stimulated amplified Faraday anomalous dispersion optical filter (SAFADOF) at cesium 1470 nm is realized. The SAFADOF is able to significantly amplify very weak laser signals and reject noise in order to obtain clean signals in strong background. Experiment results show that, for a weak signal of 50 pW, the gain factor can be larger than 25000 (44 dB) within a bandwidth as narrow as 13 MHz. Having this ability to amplify weak signals with low background contribution, the SAFADOF finds outstanding potential applications in weak signal detections.**

*OCIS codes: Filters, Faraday effect, Spectroscopy, Optical Amplifiers, Narrow bandwidth..*

The Faraday anomalous dispersion optical filter (FADOF) [1, 2] has advantages of ultra-narrow bandwidth [3], high transmittance, and high noise rejection [4, 5], which makes it an excellent frequency selection component widely used in optical signal processing [6–9] and more generally, in weak optical communication, such as free-space optical communication [10] and underwater optical communication [11]. Typically, in freespace quantum key distribution (QKD) systems [12, 13] and lidar remote sensing systems [14–17], narrow-bandwidth FADOFs are usually used to suppress out-of-band noise, thus reducing the error rate and enable observations in strong background. In such systems, the ability to extract weak signal from strong background noise relies on the narrow bandwidth of the filters, and meanwhile, the total transmission efficiency is proportional to the FADOFs' transmittance. Therefore, to enable applications in longer communication distance and higher accuracy, conventional FADOFs have been developing towards the trend of higher transmittance and narrower bandwidth.
Up to now, the FADOFs have been realized on different atomic transitions, mostly with transmittance between 40% and 100%, and

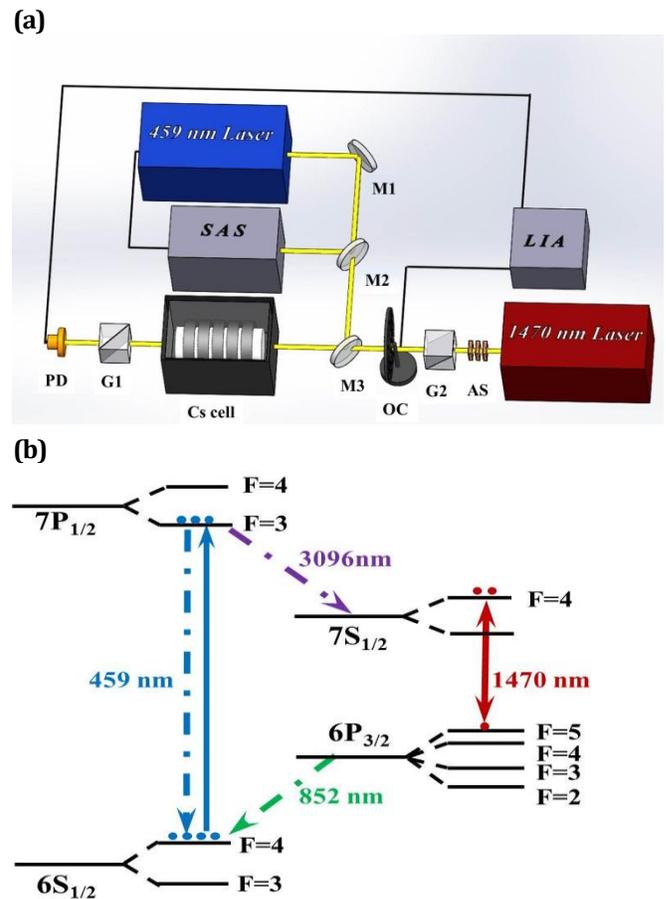

Fig. 1. (a) Experimental setup of the SAFADOF. SAS: saturated absorption spectrum. LIA: lock-in amplifier. OC: optical chopper. AS: Attenuation slices. M1: 459 nm high-reflecting mirror. M2: 459 nm partially-reflecting mirror. M3: 459 nm highreflecting and 1470 nm anti-refelcting mirror. G1 and G2: a pair of Glan-Taylor prisms whose polarization directions are orthogonal. PD: Photo diode. (b) The related energy levels of Cs atom.

equivalent noise bandwith (ENBW) around 1 GHz, such as Na 589 nm (90%, 5 GHz) [18], Rb 780 nm (83%, 2.6 GHz) [19], Rb 795 nm (70%, 1.2 GHz)[20], Cs 459 nm (98%, 1.2 GHz) [21], Cs 852 nm (88%, 0.56 GHz) [22], Cs 894 nm (77%, 0.96 GHz) [23], Sr 461 nm (63%, 1.19 GHz)

[24], etc [25–28]. An ultranarrow optical filter based on Faraday effect has been demonstrated in 2012 [29], of which the bandwidth is 6.2 MHz. However, the transmittance of this filter is only 9.7%, which finally limited its application. To break the restriction of transmittance, an atomic filter with Raman light amplification has been studied [30–32], in which a Raman light amplifier and a FADOF are used in tandem with independent Rb cells. This filter enhanced the transmittance to 85-fold compared to the case operating only with the FADOF, which expands the range of potential applications. However, for ultra weak signal detection, the amplification is still unable to meet the requirement. Also, the ability to suppress the background noise is determined by the FADOF bandwidth of 0.6 GHz, which is limited by the atomic Doppler broadening..

Here, we demonstrate a stimulated amplified Faraday anomalous dispersion optical filter (SAFADOF) at 1470 nm, which realizes the high noise rejection performance of the FADOF and the coherent amplification [35] of atomic stimulated emission simultaneously in a single Cs atomic cell. By this means an atomic filter based on population inversion is realized, and the stimulated emission process provides quite effective amplification as well as an ultra-narrow bandwidth. Experimentally, we measure a gain factor larger than 25000 (44 dB) with a probing light power of 50 pW. An ultra-narrow full width at half maximum (FWHM) of 13 MHz is achieved, and the out-band noise is totally rejected with a noise rejection ratio of 1×10^5. Being much more efficient in extracting weak signals from strong background compared with any existing atomic filters, the SAFADOF provides quite promising applications in weak signal detection in optical communication.

The experimental setup and relevant energy level structures are shown in Figure 1. A 459 nm laser stabilized to the Cs $6S_{1/2}(F=4) - 7P_{1/2}(F=3)$ transition by the saturated absorption spectrum (SAS) pumps the Cs atoms inside a 10 cm long quartz cell temperature-controlled to 410 K. After pumping, the atoms are population inverted between $7S_{1/2}(F=4)$ and $6P_{3/2}(F=5)$ states [36]. Hence with the function of the 1470 nm probing laser (coincide with the pumping laser), stimulated emission between the two states is generated, and thus the probing laser is significantly amplified. The Cs cell is placed between a pair of orthogonal Glan-Taylor prisms G1 and G2, of which the extinction ratio is $1 \times 10^5$. This also determines the out-of-band noise rejection ratio of the SAFADOF. The ring magnets outside the cell produce an axial magnetic field of about 8 Gauss, where we experimentally get the largest gain. An optical chopper together with a lock-in amplifier are used to eliminate the influences of the fluorescence generated by static superradiance [33, 37].

Due to the collective behavior of static superradiance[33, 37], population reversed atom ensemble will radiate spontaneously from the 7S1/2(F = 4) state to the 6P3/2(F = 5) state, which is much faster and stronger than that of individual atoms, and exhibit well defined direction. In our system the 1470 nm static superradiance has been observed experimentally [33]. The static superradiance light, of which the amplitude varies with the pumping power and temperature, cannot be optically filtered and will contribute to the background noise, as shown in Figure 2 (a). Such influence is eliminated by a synchronous modulation method, where the probing light is pre-modulated by an optical chopper, with a modulation frequency of 1.5 kHz. Then the detected transmission light is demodulated by a lock-in amplifier synchronized to the chopper. So that the transmission signal derived from the probing laser is well separated from the static superradiance and independently detected. Figure 2(b) illustrates the transmission spectrum before and after modulation, as well as the demodulated signal, in which the background is effectively suppressed. This method is proposed to improve the SNR of the SAFADOF, and is also applicative in other systems such as lamp-based atomic filters [24, 34], where the fluorescence has non-negligible influence.

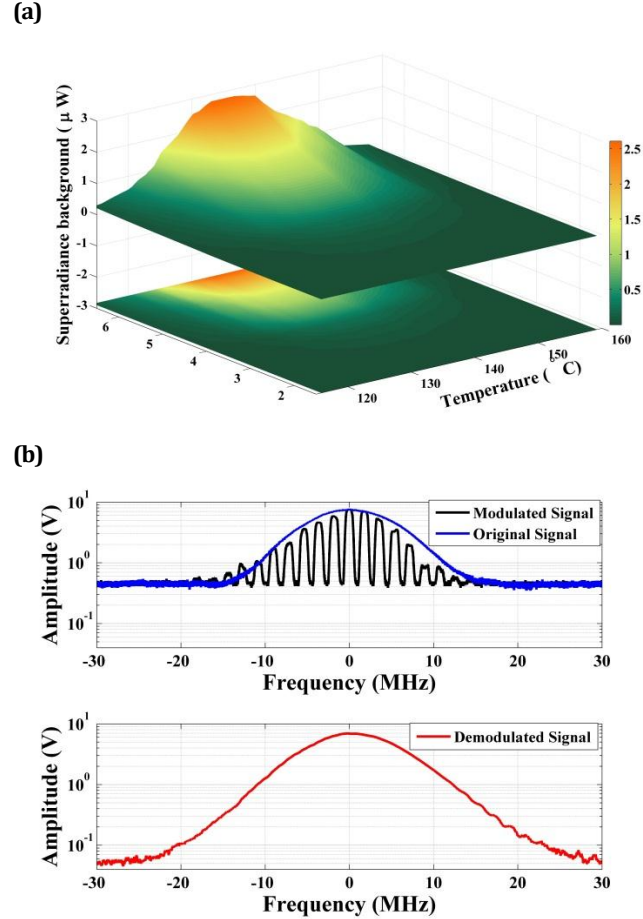

Fig. 2. (a) Detected superradiance background for various temperatures and pumping powers. (b) The transmission signal before and after modulation (up) and the demodulated signal (down).The results are obtained by scanning the laser frequency.

In the context of weak optical communication, we are interested in obtaining long communication distance and high accuracy, which requires a high transmittance of the filter to reduce the loss, or possibly, a high gain factor. Compared to the above-mentioned Raman amplified atomic filter, where the Raman gain is transformed from the coupling laser without population inversion, the SAFADOF provides much more effective amplification.

For the interaction of a two-level atomic system with a radiation field, the transition probability is given by $W(t) = |c(t)|^2$, with

$c(t) = -i \frac{\Omega}{\sqrt{\Omega^2+\Delta\omega^2}} \sin\left(\frac{\sqrt{\Omega^2+\Delta\omega^2}}{2}\right) \exp\left[-i\frac{\Delta\omega}{2}t\right]$ [3], where $\Omega$ and $\Delta\omega$ represent the Rabi frequency and the frequency detuning respectively. Thus for a radiation field on resonance, the transition probability is expressed as

$$W(t) = \sin^2(\Omega t/2) \quad (1)$$

For the atoms with average lifetime $\tau$, the distribution function of their interaction time with the radiation field is represented in the form $f(t) = \frac{1}{\tau}e^{-t/\tau}$, Then Eq. (1) transforms into:

$$\langle W \rangle = \int_0^\infty f(t)W(t)dt = \frac{1}{2}\frac{\Omega^2}{\Omega^2+\Gamma^2}, \quad (2)$$

where $\Gamma = 1/\tau$ is the spontaneous emission rate of the transition. To match our experimental conditions, considering the length of the Cs cell and the probing laser with waist w0, the variation of signal power dP during a length of dL is given by

$$dP = \frac{1}{2}\eta\Delta\rho\pi w_0^2\, h\nu \times \frac{\Omega^2}{\Omega^2+\Gamma^2}dL, \quad (3)$$

where η represents the pumping rate, which is $3.6 \times 10^6$/s in our experiment. The effective atomic density $\Delta\rho$ is $1.2 \times 10^{22}$/m3, taking

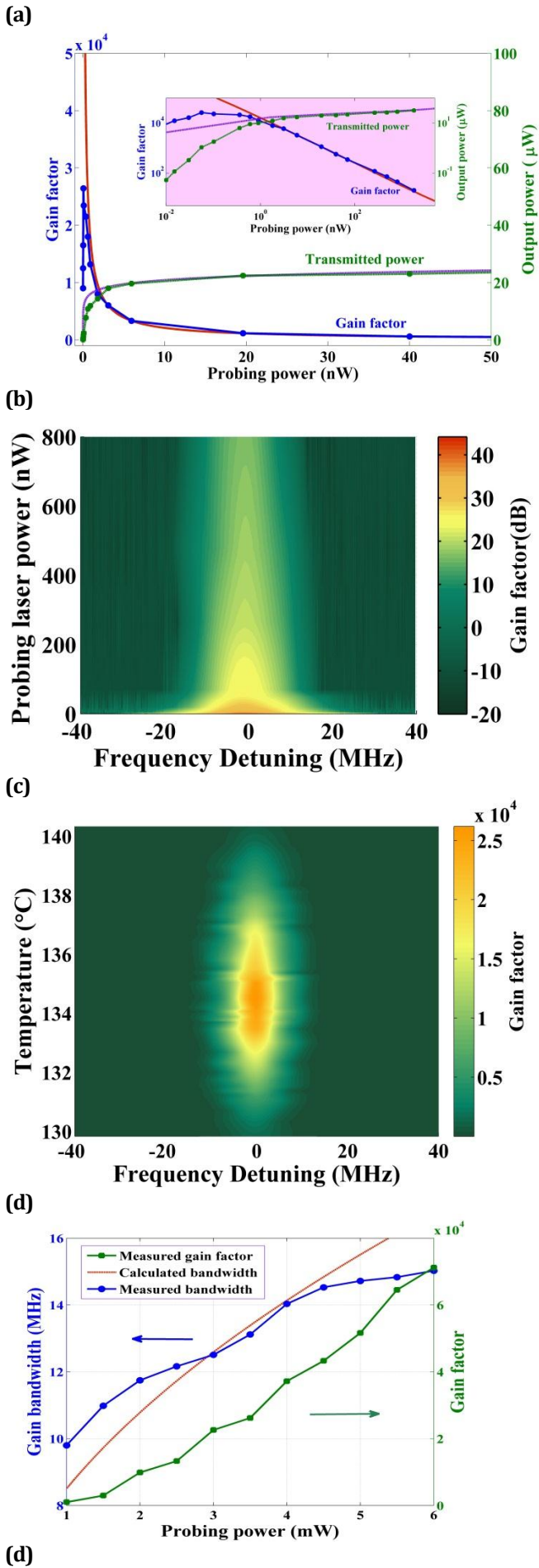

Fig. 2. (a) Calculated (red, solid) and measured (blue, solid) gain factor, as well as calculated (purple, dashed) and measured (green, dashed) transmission power of the SAFADOF for various probing powers. (b) Density plot of the gain spectrum (in dB) for various probing powers at 135 °C. (c) Gain spectrum for different temperature at the probing power of 50 pW. (d) Theoretical and experimental gain bandwidth together with the measured gain factor as a function of pumping power.

into account the atomic distribution in thermal equilibrium, the atomic density difference in the $7S_{1/2}(F=4)$ and $6P_{3/2}(F=5)$ states, and the effective atoms having Doppler-shifted frequency detuning within the linewidth of the probing laser. By integrating the expression through the interaction region L, we obtain the equation with the help of $\Omega 2 = \frac{I\Gamma 2}{2Is}$ [4]

as:

$$P - P_0 + 2\pi w_0 2 I_s \times \ln(P/P_0) = \frac{1}{2}\eta\Delta\rho\pi w_0^2 h\nu L, \qquad (4)$$

where $I_s = h\pi c\Gamma/(3\lambda^3)$ [4], is the saturation intensity, and $P_0$ is the input probing power.

Figure 3(a) displays the calculated transmitted power at resonance (purple, dashed) and gain factor (red, solid) as a function of the probing power at 135 °C with 3.5 mW 459 nm pumping power. We see that the transmitted power quickly tends to a saturation value due to the limited output capability of the atoms, thereafter the gain factor decreases in an approximate inverse proportional relationship to the probing power. Experimentally the measured transmitted power (green, dashed) and gain factor (blue, solid) are also depicted. For probing powers relatively large, the measured results agree well with the calculation, while for ultra weak probing powers the measured gain factor a sharp decline, for some optical loss that we haven't taken into consideration. The largest gain factor of more than 25000 (44 dB) is obtained at 50 pW. For various probing powers and temperatures, the gain spectrums are density plotted in Figure 3(b) and Figure 3(c) respectively. While the gain factor decreases for lower temperature due to the reduction of the Cs atomic density in the cell, for higher temperature the increased collisions between atoms decrease the coherence time of the $7S_{1/2}$ state, thus decreasing the gain factor. Such characteristics have also been reported in hydrogen maser [41], Rb and Cs atomic systems [33, 36, 42].

In the SAFADOF, the gain bandwidth is approximate to the natural linewidth of the atomic transition, for the zero-velocity selection of the atoms by Doppler-free stabilized pumping laser. However, as the power of the pumping laser increases, the saturation effect results in a Doppler broadening of Cs atoms pumped to the $7P_{1/2}$ state, and finally broadens the gain bandwidth of the SAFADOF. Considering this broadening effect, the gain bandwidth is expressed as:

$$\Delta\nu = \frac{\lambda 1}{\lambda 2} \times \Gamma\sqrt{1+s} + \Delta\nu_N \qquad (5)$$

by introducing the saturation factor $s = \frac{P/w_0}{I_s}$. $\Gamma$ represents the natural broadening of the $7P_{1/2}$ state which is 1.15 MHz, and $\Delta\nu_N$=1.82 MHz is the natural linewidth of 1470 nm transition. The first term of Eq. (5) denotes that the Doppler broadening is transmitted from the $7P_{1/2}$ state to the $7S_{1/2}$ state with $\lambda 1$ and $\lambda 2$ corresponding to the pumping and probing transitions. In our experiment the waist radius $w_0$ is 0.29 mm, and the saturation intensity $I_s$ is calculated to be 1.27 mW/cm$^2$. Figure 3 (d) shows the theoretical and experimental gain factor as well as the measured gain bandwidth depending on pumping power. We see that with the pumping power increasing, a larger gain factor is obtained. Meanwhile the gain bandwidth is broadened, which indicates that there is some optimal pumping power depending on how large the gain factor is required. The preferred pumping power will depend on the particular application.

In summary, we have experimentally investigated a SAFADOF at 1470 nm based on population inversion. The SAFADOF provides a gain factor larger than 25000 (44 dB) and an ultra-narrow bandwidth of 13 MHz, and it opens the possibility of applications in weak optical communication.

To eliminate the fluorescence background caused by superradiance of the Cs atoms, we propose a synchronous modulation method, which experimentally suppressed the background, and the method can be further expanded to other lampbased atomic filters [24, 34]. We also studied the gain factor and gain bandwidth characteristics of the SAFADOF under different probing laser powers, pumping laser powers, and temperatures. The gain factor has an approximate inverse proportional relationship with the probing power, and increases with the pumping power, while the gain bandwidth mainly increases with the pumping power. Hence a trade-off between large gain factor and narrow bandwidth must be made when determining the pumping power in practice.

**Funding sources and acknowledgments.**
National Natural Science Foundation of China(91436210, 61401036, 61531003);
Science Fund for Distinguished Young Scholars of China (61225003)..